\documentstyle[11pt]{article}

\begin{document}

\pagestyle{myheadings}
\markright{Lorente}
\flushbottom
\itemsep 0pt
\parsep 0pt
\baselineskip 12pt
\parindent 7mm
\vspace*{36pt}

\noindent {\bf QUANTUM MECHANICS ON DISCRETE SPACE \\ AND TIME}

\vskip 36pt 
\hskip 1cm{ \bf Miguel Lorente}

\vskip 12pt

\hskip 1cm {\it Departamento de F\'{\i}sica, Facultad de Ciencias}

\hskip 1cm {\it Universidad de Oviedo}

\hskip 1cm {\it E-33007, Oviedo,  Spain}

\vskip 24pt

\baselineskip 13pt
\noindent
We propose the assumption of quantum mechanics on a discrete space and time, which implies the
modification of mathematical expressions for some postulates of quantum mechanics. In particular we have
a Hilbert space where the vectors are complex functions of discrete variable. 

As a concrete example we develop a discrete analog of the one-dimensio\-nal quantum harmonic oscillator,
using the dependence of the Wigner functions in terms of Kravchuk polynomials. In this model the position
operator has a discrete spectrum given by one index of the Wigner functions, in the same way that the energy
eigenvalues are given by the other matricial index. 

Similar picture can be made for other models where the differential equation and their solutions
correspond to the continuous limit of some difference operator and orthogonal polynomial of discrete
variable.

\vskip 26pt

\noindent \parbox[t]{7mm}{\bf 1.}  \parbox[t]{115mm}{\bf INTRODUCTION}

\vskip 13pt

\indent 
In the last Symposium on Fundamental Problems in Quantum Physics [1] I explored the hypothesis of a
realistic interpretation of lattice theories based on some ontological model that presupposes some
fundamental network previous to the concept of space and time. According to this model the structure of
space-time is a consequence of the relations among these fundamental entities, and gives raise to a
discrete character of the space and time variables. 

Even the standard interpretation of quantum mechanics in our model is conserved the
assumption of a discrete space and time introduces some drastic changes in the mathematical formulation
of quantum mechanics. These consequences have been also shared by recent authors who use lattice models
as a mathematical tool. In particular non-commutative geometry leads in a natural way to difference
operators [2]. Quantum groups and q-analisis implies a deformation of space-time groups with a q-casimir
defined on a space or time lattice [3]. Lattice field theories are widely spread in an attempt to
overcome infinities in perturbation theories [4]. Special functions and orthogonal polynomials on
continuous variable are studied together with those of discrete variable [5]. Non-euclidean
crystallography requires the use of discrete groups of hyperbolie type that has been recently
developped [6]. Recent literature advocates for the application of discrete models to unify general
relativity and quantum mechanics as Wheeler, Ponzano and Regge and others have proposed [7]. For this
purpose some modern tools are used, such as discrete topology and partition functions defined over
symplicial networks.

\vskip 26pt

\noindent \parbox[t]{7mm}{\bf 2.}  \parbox[t]{117mm}{\bf THE POSTULATES OF QUANTUM MECHANICS ON $Z^4$}

\vskip 13pt

\indent
The assumption of discrete space and time imposses some restrictions on the mathematical expressions
for the postulates of Quantum Mechanics. The question is now whether this assumption keeps the analogy
with standard formulation in the continuum case (in the limit both formulations coincide) and at the
same time avoids the unwanted infinities. The answer is, generally speaking, in the affirmative. 

In order to be more explicit, we start with the Hilbert space: this must be defined over the gaussian
numbers (complex of integer components) with scalar products constructed with summation instead of
integrals
$$\sum\limits_{j=a}^{b} f_{j}^{*} g_j$$
where we use the notation $f_j\equiv f\left(j\varepsilon \right)$, $\varepsilon$ being the
fundamental length of the one-dimensional lattice, $j$ integer numbers.

As an example, we take the following orthonormal basis

$$f_j\left({{k}_{m}}\right)={\frac{1}{\sqrt {N}}}{\left({{\frac{1+{\frac{1}{2}}i\varepsilon
{k}_{m}}{1-{\frac{1}{2}}i\varepsilon {k}_{m}}}}\right)}^{j}\ \ \ ,\ \ \
j=0,1,\cdots N-1$$
where

$${k}_{m}\equiv {\frac{2}{\varepsilon }}{\rm tg}\ {\frac{\pi m}{N}}\ \ \ ,\ \ \
m=0,1,\cdots N-1$$
satisfying 

$$\sum\limits_{j\ =\ 0}^{N-1} {f}_{j}^{{}^*}\left({{k}_{m}}\right){f}_{j}\left({{
k}_{m'}}\right) ={\delta }_{mm'}$$
with respect to which a finite Fourier transform (equivalent to a Fourier series) can be defined
$${F}_{j}=\sum\limits_{m\ =\ 0}^{N-1} {a}_{m}{f}_{j}\left({{
k}_{m}}\right)\ \ \ ,\ \ \ {a}_{m}=\sum\limits_{j\ =\ 0}^{N-1} {f}_{
j}^{{}^*}\left({{k}_{m}}\right){F}_{j}$$

Notice that although the space-time variables are discrete the functions can be continuous.

If we consider observables, the corresponding operators must be self adjoint with respect to the scalar
product mentioned before, and the spectrum is always discrete.  As an example, we give the eigenvalues
and eigenfunctions of the position and momentum operators.
\[X:\ \ \ j{\delta }_{j\ell }=\ell {\delta }_{j\ell }\ ,\ \ \ell \ {\rm fix}\ ;\ \ j=0,1,\cdots N-1\]
\[P:\ \ \ -i{\frac{1}{\varepsilon }}{\Delta }_{j}
{f}_{j}\left({{k}_{m}}\right)={\frac{{k}_{m}}{1-{\frac{1}{2}}\varepsilon {k}_{\rm
m}}}{f}_{j}\left({{k}_{m}}\right)\]
where ${\delta }_{j\ell }$ is the Kronecker function and ${f}_{j}\left({{k}_{m}}\right)$ is defined as
before.

With the help of the scalar product we can defined espectation values, projection operators, density
matrix, uncertainties or mean values of some operators.

Suppose now that a physical system is represented by a state vector $\psi \left({t}\right)$
depending on discrete time $\left({t=n\tau }\right)$. If $H$ is the operator corresponding to the
hamiltonian of the system (for simplicity we take
$H$ constant in time) the Schr\"odinger equation for that system is 
$${\frac{i}{\tau }}{\Delta }_{n}{\psi }_{n}=H{\tilde{\Delta }}_{n}{\psi }_{n}$$
the solution of which is
$${\psi }_{n}={\left({{\frac{1-{\frac{1}{2}}i\tau H}{1+{\frac{1}{2}}i\tau H}}}\right)}^{n}{\psi
}_{0}$$
with the initial condition ${\psi }_{0}$.

Here ${\Delta }_{n}$ is the forward difference operator ${\Delta }_{n}{\psi }_{n}={\psi }_{n+1}-{\psi
}_{n}$
 and ${\tilde{\Delta }}_{n}$ the mean operator ${\tilde{\Delta }}_{n}{\psi }_{n}={\frac{1}{2}}\left({{\psi
}_{n+1}+{\psi }_{n}}\right)$

As in the continuous case we can define an unitary evolution operator
$${U}_{n}\equiv {\left({{\frac{1-{\frac{1}{2}}i\tau H}{1+{\frac{1}{2}}i\tau H}}}\right)}^{n}$$
which is unitary (because $H$ is self adjoint) and satisfies the difference equation
$${\frac{i}{\tau }}{\Delta }_{n}{U}_{n}=H{\tilde{\Delta }}_{n}{U}_{n}$$

If we use the Heisenberg picture the evolution of some operators in time is given by

\begin{equation}
{A}_{n}={U}_{n}^{+}{A}_{0}{U}_{n}
\end{equation}
where $A_n$ is some operators depending on discrete time and $U_n$ is the evolution operator defined
before. The Heisenberg equation now reads [8]

$${\frac{i}{\tau }}\Delta {A}_{n}={\frac{1}{1-{\frac{1}{2}}i\tau
H}}\left[{A_n,H_n}\right]{\frac{1}{1+{\frac{1}{2}}i\tau H}}$$
the solution of which is $(1)$.

An other scheme may be used if we take the symmetric difference operator

$${\delta }_{n}{A}_{n}\equiv \left({{A}_{n+{\frac{1}{2}}}-{A}_{n-{\frac{1}{2}}}}\right),$$
then

\[{\frac{i}{\tau }}{\delta }_{n}{A}_{n}={\frac{1}{{\left({1+{\frac{1}{4}}{\tau
}^{2}{H}^{2}}\right)}^{{\frac{1}{2}}}}}\left[{A_n,H_n}\right]{\frac{1}{{\left({1+{\frac{1}{4}}{\tau
}^{2}{H}^{2}}\right)}^{{\frac{1}{2}}}}}\]

Also we have

\[{\frac{i}{\tau }}\left({{A}_{n+1}-{A}_{n-1}}\right)={\frac{1}{1+{\frac{1}{4}}{\tau
}^{2}{H}^{2}}}2\left({1-{\frac{{\tau
}^{2}}{4}}}\right)\left[{A_n,H_n}\right]{\frac{1}{1+{\frac{1}{4}}{\tau }^{2}{H}^{2}}}\]

Some simplification can be achieved if we take the particular case $H^2=1$. It can be easily proved the
following equation for the operator in the Heisenberg picture:

\[
\begin{array}{l}
{\frac{i}{\tau }}{\Delta }_{n}{A}_{n}={\frac{1}{\left({1+{\frac{{\tau
}^{2}}{4}}}\right)}}\left({{\frac{1-{\frac{i}{2}}\tau H}{1+{\frac{i}{2}}\tau
H}}}\right)\left[{{A}_{n},H}\right] \bigskip\\
{\frac{i}{\tau }}{\nabla }_{n}{A}_{n}={\frac{1}{\left({1+{\frac{{\tau
}^{2}}{4}}}\right)}}\left({{\frac{1+{\frac{i}{2}}\tau H}{1-{\frac{i}{2}}\tau
H}}}\right)\left[{{A}_{n},H}\right] \bigskip\\
{\left({{\frac{i}{\tau }}}\right)}^{2}{\Delta }_{n}{\nabla
}_{n}{A}_{n}={\frac{1}{{\left({1+{\frac{{\tau
}^{2}}{4}}}\right)}^{2}}}\left[{\left[{{A}_{n},H}\right],\ H}\right] \bigskip\\
{\frac{i}{\tau }}{\delta }_{n}{A}_{n}={\frac{1}{{\left({1+{\frac{{\tau
}^{2}}{4}}}\right)}^{2}}}\left[{{A}_{n},H}\right] \bigskip\\
{\frac{i}{\tau }}\left({{A}_{n+1}-{A}_{n-1}}\right)={\frac{2\left({1-{\frac{{\tau
}^{2}}{4}}}\right)}{{\left({1+{\frac{{\tau }^{2}}{4}}}\right)}^{2}}}\left[{{A}_{n},H}\right]
\end{array}\]

In the last three equations the dependence on the Hamiltonian operator $H$ is linear as in the
continuous case. 

The realization of operators in the coordinate or position representation is given by the substitution 

$$X\rightarrow j\varepsilon ,\ \ \ \ \ P\rightarrow -{\frac{i}{\varepsilon }}{\Delta }_{j}$$

The substitution is not unique. We will discuss in the next section some different realization for the
position and momentum operator of he harmonic oscillator.

\vskip 26pt

\noindent \parbox[t]{7mm}{\bf 3.}  \parbox[t]{117mm}{\bf QUANTUM HARMONIC OSCILLATOR OF DISCRETE VARIABLE}

\vskip 13pt

\indent
The quantum harmonic oscillator is described by the Schr\"odinger equation

$$
{\frac{\hbar \omega }{2}}\left[{-{\frac{{d}^{2}}{d{\xi }^{2}}}+{\xi
}^{2}}\right]\psi\left({\xi}\right)=\lambda \psi\left({\xi}\right)
$$
with $\omega \equiv \sqrt {{\frac{k}{M}}},\ \ \ \ \ \xi =\alpha s,\ \ \ \ \ \alpha \equiv \sqrt
{{\frac{M\omega }{\hbar }}},\ \ \ \ \ \lambda ={\frac{2E}{\hbar \omega }}$

For simplicity, we take $\alpha = 1$

The normalized solutions are

\begin{equation}
{\psi }_{n}\left({s}\right)={\left({{\pi
}^{{\frac{1}{2}}}{2}^{n}n!}\right)}^{-{\frac{1}{2}}}{e}^{-{s}^{2}/2}{H}_{n}\left({s}\right)\ \ \ \ \ ,\
\ \ \ \ n=0,1,2,\cdots 
\end{equation}
where ${H}_{n}\left({s}\right)$ are the Hermite polynomials.

The ${\psi }_{n}\left({s}\right)$ are eigenfunctions corresponding to the eigenvalues $\lambda =2n+1$,
and they satisfy the following recurrence relations: 

\begin{eqnarray}
{\rm i)} & & 2s{\psi }_{n}\left({s}\right)=\sqrt {2\left({n+1}\right)}\ {\psi
}_{n+1}\left({s}\right)+\sqrt {2n}\ {\psi }_{n-1}\left({s}\right) \bigskip \\
{\rm ii)} & & 2{\frac{d}{ds}}{\psi }_{n}\left({s}\right)=-\sqrt {2\left({n+1}\right)}\ {\psi
}_{n+1}\left({s}\right)+\sqrt {2n}\ {\psi }_{n-1}\left({s}\right)
\end{eqnarray}

From these two relations one defines the creation and annihilation operators:

\begin{eqnarray}
{a}^{\dagger}{\psi }_{n}\left({s}\right)&\equiv & {\frac{1}{\sqrt
{2}}}\left({s-{\frac{d\,}{d\,s}}}\right){\psi }_{n}\left({s}\right)=\sqrt {n+1}{\psi
}_{n+1}\left({s}\right) \bigskip \\ a{\psi }_{n}\left({s}\right) &\equiv & {\frac{1}{\sqrt
{2}}}\left({s+{\frac{d\,}{d\,s}}}\right){\psi }_{n}\left({s}\right)=\sqrt {n}{\psi
}_{n-1}\left({s}\right)
\end{eqnarray}

It is well known that the Hermite polynomials are the continuous limit of the Kravchuk polynomials of
discrete variables ${k}_{n}\left({x}\right)$ and the weight function of the Hermite polynomials is the
continuous limit of the binomial distribution $\rho \left({x}\right)$ which in turns is the weight
function of the Kravchuk polynomial [9]. But the product of the Kravchuk polynomials time their weight
function is proportional to the Wigner functions ${d}_{mm'}^{j}\left({\beta }\right)$ that appear in the
generalized spherical functions 

\begin{equation}
{D}_{mm'}^{j}\left({\alpha ,\beta ,\gamma }\right)=\exp\ \left({-im\alpha
}\right){d}_{mm'}^{j}\left({\beta }\right)\exp\ \left({-im'\gamma }\right),
\end{equation}
namely,

\begin{equation}
{d}_{mm'}^{j}\left({\beta }\right)={\left({-1}\right)}^{m-m'}{\frac{\sqrt
{\rho\left({x}\right)}}{{d}_{n}}}{k}_{n}^{\left({p}\right)}\left({x,N}\right)
\end{equation}
where $d_n$ is some normalization constant, and $n=j-m\ ,\ \ \ x=j-m'\ ,\ \ \ p={\sin}^{2}\ \left({\beta
/2}\right)$.
 
This connection betwen the functions of discrete and continuous variable suggests that the solution of
the quantum harmonic oscillator are the continuous limit, up to a factor, of the Wigner functions. In
order to prove this Ansatz we compare the recurrence relations of the two types of functions as it is
done for the orthogonal polynomials of discrete variable. In our case we take the differential equation
for the Wigner function [10] 

\[\pm {\frac{d}{d\beta}}\ {d}_{mm'}^{j}\left({\beta }\right) +{\frac{m'-m\;{\rm
cos}\;\beta}{\sin\ \beta }}\ {d}_{mm'}^{j}\left({\beta }\right) =
\]\vspace{-1,5ex}
$$=\sqrt{\left({j\mp m}\right)\left({j\pm m+1}\right)}\ {d}_{m\pm 1,m'}^{j}\left({\beta
}\right)
\eqno(9)(10)$$

\setcounter{equation}{10}

From this we deduce two recurrence relationes: 
\begin{eqnarray} 
{\rm i)}& &2\ {\frac{m'-mcos\ \beta }{\sin\ \beta }}{d}_{mm'}^{j}\left({\beta
}\right)=\nonumber \\  
\noalign{\smallskip}& &\hspace{2cm}=\sqrt {\left({j-m}\right)\left({j+m+1}\right)}{d}_{m+1,m'}^{j}\left({\beta
}\right)+ \nonumber \\  
\noalign{\smallskip}& &\hspace{3cm} + \sqrt
{\left({j+m}\right)\left({j-m+1}\right)}{d}_{m-1,m'}^{j}\left({\beta }\right)
\\ 
\noalign{\bigskip}{\rm ii)} & &\sqrt
{\left({j+m'}\right)\left({j-m'+1}\right)}{d}_{m,m'-1}^{j}\left({\beta }\right)- \nonumber \\
\noalign{\smallskip}& &\hspace{3cm}-\sqrt
{\left({j-m'}\right)\left({j+m'+1}\right)}{d}_{m,m'+1}^{j}\left({\beta }\right)= \bigskip
\nonumber \\  
\noalign{\smallskip}& &\hspace{2cm}= \sqrt{\left({j-m}\right)\left({j+m+1}\right)}{d}_{m+1,m'}^{j}-
\nonumber \\
\noalign{\smallskip}& &\hspace{3cm}-\sqrt {\left({j+m}\right)\left({j-m+1}\right)}{d}_{m-1,m'}^{j}
\end{eqnarray}

Note that the last expression has been obtained with the help of the well known property of Wigner functions

$$
{d}_{mm'}^{j}\left({\beta }\right)={\left({-1}\right)}^{m-m'}{d}_{m'm}^{j}\left({\beta
}\right)
$$

We suppose that 

\begin{equation}
\lim\limits_{N\ \rightarrow \ \infty }^{}\
{C}_{n}\left({N}\right){d}_{mm'}^{j}\left({\beta }\right)={\psi }_{n}\left({s}\right)
\end{equation}
where we take $m=j-n$, $m'=j-x$, $N=2j$, $x=Np + \sqrt {2Npq}\ s$, and
${C}_{n}\left({N}\right)$ some normalization constant to be determined.

We compare the recurrence relations i) that is to say, formulas (3) and (11). We divide the second one by
$\sqrt {j}$ and substitute ${d}_{mm'}^{j}\left({\beta }\right)$ by
${\frac{{v}_{n}\left({x}\right)}{{C}_{n}\left({N}\right)}}$, with ${v}_{n}\left({x}\right)\equiv
{C}_{n}\left({N}\right){d}_{j-n,j-x}^{j}\left({\beta }\right)$.
 
The result is:

\begin{eqnarray*}
2{\frac{j-x-\left({j-n}\right)\cos\ \beta }{\sqrt {j}\sin\ \beta
}}{\frac{{v}_{n}\left({x}\right)}{{C}_{n}\left({N}\right)}}&=&\sqrt
{{\frac{2n\left({N-n+1}\right)}{2j}}}{\frac{{v}_{n-1}\left({x}\right)}{{C}_{n-1}\left({x}\right)}}\ + \\
\noalign{\smallskip}&+&\sqrt
{{\frac{2\left({N-n}\right)\left({n+1}\right)}{2j}}}{\frac{{v}_{n+1}\left({x}\right)}{{C}_{n+1}
\left({x}\right)}}
\end{eqnarray*}
or

\begin{eqnarray*}
2\left({s+{\frac{\left({2p-1}\right)n}{\sqrt
{2Npq}}}}\right){\frac{{v}_{n}\left({x}\right)}{{C}_{n}\left({N}\right)}}&=&\sqrt
{2\left({n+1}\right)}\sqrt
{1-{\frac{n}{N}}}{\frac{{v}_{n+1}\left({x}\right)}{{C}_{n+1}\left({N}\right)}}+ \\
\noalign{\smallskip}&+&\sqrt {2n}\sqrt
{1-{\frac{n-1}{N}}}{\frac{{v}_{n-1}\left({x}\right)}{{C}_{n}\left({N}\right)}}
\end{eqnarray*}

In the limit $N\rightarrow \infty $ this expression goes to the recurrence relation (3) provided 
${\frac{{C}_{n}\left({N}\right)}{{C}_{n+1}\left({N}\right)}}={\frac{{C}_{n}\left({N}\right)}
{{C}_{n-1}\left({N}\right)}}=1$, or ${C}_{n}\left({N}\right)= {\rm const}=1$.

The recurrence relations ii) formules (4) and (12) can be compared by the same method. We substitute

\begin{equation}
{v}_{n}\left({x}\right)\equiv{d}_{j-n,\ j-x}^{j}\left({\beta }\right)
\end{equation}
in (12) and divide both sides by $\sqrt {N}$.

The result is

\begin{eqnarray*}
\lefteqn{\sqrt
{{\frac{\left({N-x}\right)}{N}}\left({x+1}\right)}{v}_{n}\left({x+1}\right)-\sqrt
{x{\frac{\left({N-x+1}\right)}{N}}}{v}_{n}\left({x-1}\right)= }\bigskip \\
&=& \sqrt
{n{\frac{\left({N-n+1}\right)}{N}}}{v}_{n-1}\left({x}\right)-\sqrt
{{\frac{\left({N-n}\right)}{N}}\left({n+1}\right)}{v}_{n+1}\left({x}\right)
\end{eqnarray*}

Substituting $x=Np+\sqrt {2Npq}s$, and extracting $\sqrt {2Np}\equiv {\frac{1}{h}}$ in the left side, we
obtain

\begin{eqnarray*}
\lefteqn{{\frac{1}{h}}\left\{{\sqrt {\left({1-{\frac{x}{N}}}\right)\left({1+\sqrt
{{\frac{2q}{Np}}}s+{\frac{1}{Np}}}\right)}{v}_{n}\left({x+1}\right) -}\right. }\bigskip \\
& &\left. -\sqrt {\left({1+\sqrt{{\frac{2q}{Np}}}s}\right)\left({1-{\frac{x-1}{N}}}\right)}
{v}_{n}\left({x-1}\right)\right\}=\bigskip \\
&=& \sqrt
{2n\left({1-{\frac{n-1}{N}}}\right)}{v}_{n-1}\left({x}\right)-\sqrt
{2\left({n+1}\right)\left({1-{\frac{n}{N}}}\right)}{v}_{n+1}\left({x}\right)
\end{eqnarray*}

In the limit $N\rightarrow \infty$, $h\rightarrow 0$ this expression goes to

\[\lim\limits_{h\ \rightarrow \ 0}^{}\ {\frac{1}{h}}\left\{{{\psi }_{n}\left({s+h}\right)-{\psi
}_{n}\left({s-h}\right)}\right\}=-\sqrt {2\left({n+1}\right)}{\psi }_{n}\left({s}\right)+\sqrt {2n}{\psi
}_{n-1}\left({s}\right)\]
that coincides with (4)

We can use these results to construct creation and annihilation operators for the Wigner functions. We
define

\begin{eqnarray}
A{d}_{mm'}^{j}\left({\beta }\right) &\equiv & \frac{1}{\sqrt {2}} \left\{  \frac{m'-mcos\
\beta }{\sqrt {j}\sin\ \beta }{d}_{mm'}^{j}\left({\beta }\right)+ \right. \nonumber\\
& &\left. +\sqrt
{\frac{\left({j+m'}\right)\left({j-m'+1}\right)}{4j}}{d}_{m,m'-1}^{j}\left({\beta
}\right)- \right. \nonumber\\
& &\left. -\sqrt
{\frac{\left({j-m'}\right)\left({j+m'+1}\right)}{4j}}{d}_{m,m'+1}^{j}\left({\beta}\right)\right\}=
 \nonumber\\
& &=\sqrt {{\frac{\left({j-m}\right)\left({j+m+1}\right)}{2j}}}{d}_{m+1,m'}^{j}\left({\beta }\right)
 \bigskip \\
{A}^\dagger {d}_{mm'}^{j}\left({\beta }\right)&\equiv &\frac{1}{\sqrt {2}}\left\{\frac{m'-mcos\
\beta }{\sqrt {j}\sin\ \beta }{d}_{mm'}^{j}\left({\beta }\right) - \right. \nonumber\\
& &\left. -\sqrt
{\frac{\left({j+m'}\right)\left({j-m'+1}\right)}{4j}}{d}_{m,m'-1}^{j}\left({\beta
}\right) + \right. \nonumber\\
& & \left. +\sqrt
{\frac{\left({j-m'}\right)\left({j+m'+1}\right)}{4j}}{d}_{m,m'+1}^{j}\left({\beta
}\right)\right\}= \nonumber\\
 & &=\sqrt
{{\frac{\left({j+m}\right)\left({j-m+1}\right)}{2j}}}{d}_{m-1,m'}^{j}\left({\beta
}\right)\end{eqnarray}

Using the limit of the recurrence relations we obtain

\begin{equation}
\lim\limits_{N\ \rightarrow \ \infty }^{}\ A{d}_{mm'}^{j}\left({\beta }\right)={\frac{1}{\sqrt
{2}}}\left({s+{\frac{\mit d\,}{\mit d\,s}}}\right){\psi }_{n}\left({s}\right)\equiv a{\psi
}_{n}\left({s}\right)
\end{equation}
\begin{equation}
\lim\limits_{N\ \rightarrow \ \infty }^{}\ {A}^\dagger {d}_{mm'}^{j}\left({\beta }\right)={\frac{1}{\sqrt
{2}}}\left({s-{\frac{\mit d\,}{\mit d\,s}}}\right){\psi }_{n}\left({s}\right)\equiv {a}^\dagger {\psi
}_{n}\left({s}\right)
\end{equation}

Relations (15) and (16) suggest that the creation and annihilation operators are connected with the
raising  and lowering operators for the spherical harmonics ${Y}_{jm}$. In fact, multiplying (15) and
(16) by ${Y}_{jm'}$ and adding for $m'$ we have

$$A\sum\limits_{m'}^{} {d}_{mm'}^{j}\left({\beta }\right){Y}_{jm'}=\sqrt
{{\frac{\left({j-m}\right)\left({j+m+1}\right)}{2j}}}\sum\limits_{m'}^{} {d}_{m+1,m'}^{j}\left({\beta
}\right){Y}_{jm'}$$
or

\begin{equation}
A{Y}_{jm}={\frac{1}{\sqrt {2j}}}\sqrt
{\left({j-m}\right)\left({j+m+1}\right)}{Y}_{j,m+1}={\frac{1}{\sqrt {2j}}}J_{+}{Y}_{jm}
\end{equation}
similarly,

$${A}^{\dagger}\sum\limits_{}^{} {d}_{mm'}^{j}\left({\beta }\right){Y}_{jm'}=\sqrt
{{\frac{\left({j+m}\right)\left({j-m+1}\right)}{2j}}}\sum\limits_{m'}^{} {d}_{m-1,m'}^{j}\left({\beta
}\right){Y}_{jm'}$$
or 

\begin{equation}
{A}^{\dagger}{Y}_{jm}={\frac{1}{\sqrt {2j}}}\sqrt
{\left({j+m}\right)\left({j-m+1}\right)}{Y}_{j,m-1}={\frac{1}{\sqrt {2j}}}J_{-}{Y}_{jm}
\end{equation}

In order to make more transparent the connection between the creation and annihilation operators with
the raising and lowering operators of the spherical harmonics, we take the commutation and
anticommutation relations of the former operators.

\begin{eqnarray*}
\left({A{A}^{\dagger}-{A}^{\dagger}A}\right){Y}_{jm}&=&{\frac{1}{2j}}
\left({{J}_{+}{J}_{-}-{J}_{-}{J}_{+}}\right){Y}_{jm}= \\ 
&=&{\frac{1}{2j}}
2{J}_{z}{Y}_{jm}={\frac{m}{j}}{Y}_{jm}=\left({1-{\frac{n}{j}}}\right){Y}_{jm}
\end{eqnarray*}

Substituting ${Y}_{jm}=\sum\limits_{m'}^{} {d}_{mm'}^{j}\left({\beta }\right){Y}_{jm'}$ we get

\begin{equation}
\left[{A,{A}^{\dagger}}\right]{d}_{mm'}^{j}\left({\beta
}\right)=\left({1-{\frac{n}{j}}}\right){d}_{mm'}^{j}\left({\beta }\right)
\end{equation}
which in the limit $j\ \rightarrow \ \infty$ goes to

$$\left[{a,{a}^{\dagger}}\right]{\psi }_{n}\left({s}\right)={\psi }_{n}\left({s}\right)$$

Similarly 

\begin{eqnarray*}
\left({A{A}^{\dagger}+{A}^{\dagger}A}\right){Y}_{jm}
&=&{\frac{1}{2j}}\left({{J}_{+}{J}_{-}+{J}_{-}{J}_{+}}\right){Y}_{jm}=
{\frac{1}{j}}\left({{\vec{J}}^{2}-{J}_{z}^{2}}\right){Y}_{jm}= \bigskip \\
& = &{\frac{1}{j}}\left({j\left({j+1}\right)-{m}^{2}}\right){Y}_{jm}=
\left\{{\left({2n+1}\right)-{\frac{{n}^{2}}{j}}}\right\}{Y}_{jm}
\end{eqnarray*}
or

\begin{equation}
\left({A{A}^{\dagger}+{A}^{\dagger}A}\right){d}_{mm'}^{j}\left({\beta
}\right)=\left\{{\left({2n+1}\right)-{\frac{{n}^{2}}{j}}}\right\}{d}_{mm'}^{j}\left({\beta }\right)
\end{equation}
which in the limit $j\rightarrow \infty$ goes to 

$$\left({a{a}^{\dagger}+{a}^{\dagger}a}\right){\psi }_{n}\left({s}\right)=\left({2n+1}\right){\psi
}_{n}\left({s}\right)$$

If we multiply both sides by $\hbar \omega /2$ we obtain the eigenvalue equation for the hamiltonian.

The interpretation of this model can be taken from the quantum harmonic oscillator of continuous variable. The
energy levels are equally distant by the amount $\hbar \omega$ and are labelled by $n=0,1,2,\cdots
\infty$. In the quantum harmonic oscillator of discrete variable we have also the discrete eigenvalues of
the hamiltonian connected with the index $m=j-n$ of the Wigner function ${d}_{mm'}^{j}\left({\beta
}\right)$. These values are equally separeted but finite $\left({m=-j,\cdots +j}\right)$. 

Similarly the eigenvalue of the position operator $A+A^+$ are also discrete and connected to the index $m'=j-x$
of the Wigner functions but finite $\left({m'=-j,\cdots ,+j}\right)$. 

The integer numbers $x=0,1,\cdots 2j$ are related to the quantity $x=\alpha s$ where $s$ is the continuous
variable and $\alpha =\sqrt {M\omega /\hbar }$. Since $x$ is a pure number and $s$ has the dimension of a
length, the spacing of the one-dimensional lattice is equal to $1/\alpha =\sqrt {\hbar /M\omega }$.
Therefore the Planck's constant $\hbar$ play an role with respect to discrete space similar to the role with
respect to discrete energy values.

\vskip 26pt

\noindent \parbox[t]{7mm}{\bf 24.}  \parbox[t]{117mm}{\bf CONCLUDING REMARKS}

\vskip 13pt

\indent
The analysis we have made for the quantum harmonic oscillator of discrete variable can be applied to
other cases, where the functions involved are orthogonal polynomials of continuous variable the limit of
which are some orthogonal polynomial of discrete variables; we give some examples:

1. The function ${f}_{j}\left({{k}_{m}}\right)$ described in section 2 are polynomials of discrete variable
the continuous limit of which are the exponential function. We have developed a new scheme for the
Klein-Gordon and Dirac field equation that can be extended to lattice gauge theories [11].

2. The solution of the Schr\"odinger equation for the hidrogen atom is given in term of the orthogonal
Laguerre polynomials and the spherical harmonics. The radical equation can be translated into the
difference equation for the Meixner polynomial of discrete variable. 

3.	The quantification of the electromagnetic fiels leads to the D'Alam\-bert equation the solution of
which are given in terms of the Bessel spherical functions that are related to the trigonometric functions.
These functions suggest the connection with the orthogonal polynomials of discrete variable, that are
solutions of difference equations of the hypergeometric type. General speaking a parallel study of
discrete and continuous model can be made similar to that made by the russian school of mathematicians
with respect to orthogonal polynomials.

\vskip 26pt

\noindent \parbox[t]{117mm}{\bf ACKNOWLEDGEMENTS}

\vskip 13pt

\indent
I would like to thank the organizers for the invitation to this meeting.

I want to thank also to Prof. Andr\'e Ronveaux for very illuminating suggestions about the connection
between orthogonal polynomials of discrete and continuous variable and Prof. A. Galindo, A. Fz. Ra\~nada,
R. Alvarez Estrada for very interesting comments about this model. This work has been supported partially by
D.G.I.C.Y.T. (grant PB94-1318).

\end{document}